\documentclass[twocolumn,aps,showpacs,prl]{revtex4}
\usepackage{graphicx}
\usepackage{epstopdf}

\begin{document}

\title{Iron-based layered superconductor LaO$_{1-x}$F$_x$FeAs: an
antiferromagnetic semimetal}
\author{Fengjie Ma$^{1}$}
\author{Zhong-Yi Lu$^{2}$}

\email{zlu@ruc.edu.cn}

\date{\today}

\affiliation{$^{1}$Institute of Theoretical Physics, Chinese Academy
of Sciences, Beijing 100080, China }

\affiliation{$^{2}$Department ofPhysics, Renmin University of China,
Beijing 100872, China}

\begin{abstract}
We have studied the newly found superconductor compound
LaO$_{1-x}$F$_x$FeAs through the first-principles density functional
theory calculations. We find that the parent compound LaOFeAs is a
quasi-2-dimensional antiferromgnetic semimetal with most carriers
being electrons and with a magnetic moment of $2.3\mu_B$ located
around each Fe atom on the Fe-Fe square lattice. Furthermore this is
a commensurate antiferromagnetic spin density wave due to the Fermi
surface nesting, which is robust against the F-doping. The observed
superconduction happens on the Fe-Fe antiferromagnetic layer,
suggesting a new superconductivity mechanism, mediated by the spin
fluctuations. An abrupt change on the Hall measurement is further
predicted for the parent compound LaOFeAs.
\end{abstract}


\maketitle


Just recently an iron-based layered compound LaOFeAs was reported to
show superconductivity after doping F atoms to replace O atoms at a
concentration of 3-13 atom\%, with the highest critical temperature
of about 26K\cite{kamihara}. In comparison with those high T$_c$
superconductors which superconduction happens on CuO$_2$
layers\cite{muller}, the compound LaOFeAs possesses conduction at
iron-based FeAs layers and further goes superconducting after doping
F atoms. This strongly suggests that there be a new
superconductivity mechanism underlying because unlike the case of
superconduction at CuO$_2$ layers it is the transition metal atoms,
especially here the Fe atoms, that directly play a role in
conduction and superconduction, in which there would be many
interactions and degrees involved, like electronic and magnetic.

Clearly, the understanding of electronic and magnetic structures of
the parent compound LaOFeAs and its F-doped derivatives is the key
basis to understanding the new possible superconductivity mechanism
underlying. There have now been a number of theoretical works with
target of understanding electronic structure of the compound LaOFeAs
\cite{singh,xu,kotliar}. All of these works find that the compound
LaOFeAs is a nonmagnetic metal with a low density of carriers and
its normal phase is located at the borderline of magnetic phases.
The work \cite{kotliar} further suggests that the correlated effect
would be considered.

It is well-known that at low temperatures in the order of $3d$
transition metals in their element form, Cr and Mn usually take
antiferrogametic state while Co and Ni are ferromagnetic, the
in-between Fe displays a variety of magnetic phases, for example,
bcc-Fe is a strong ferromagnet, but fcc-Fe may be nonmagnetic,
ferromagnetic, antiferromagnetic, or spin density wave, dependent
upon the local environment and the lattice constant \cite{moruzzi}.
Accordingly, we expect that the Fe atoms would show novelty in
LaOFeAs compound. Indeed our first-principles calculations show very
interesting finding that the compound LaOFeAs is an
antiferromagnetic (AFM) semimetal rather than nonmagnetic (NM) metal
due to the Fermi surface nesting.

We performed the first-principles density functional theory
electronic structure calculations by using the plane wave basis
method\cite{pwscf}. In the calculations, we adopted the local (spin)
density approximation (L(S)DA) and the generalized gradient
approximation (GGA) of Perdew-Burke-Ernzerhof (PBE)\cite{pbe} for
the exchange-correlation potentials. And the ultrasoft
pseudopotentials \cite{vanderbilt} were used to model the
electron-ion interactions. After the full convergence test, the
kinetic energy cut-off and the charge density cut-off of the plane
wave basis were chosen to be 800eV and 4800eV, respectively. The
Gaussian broadening technique was used and a mesh of $36\times
36\times 18$ k-points were sampled for the Brillouin-zone
integration. For all relaxed structures, the convergence sets by the
forces being smaller than 0.01eV/\AA.

The parent compound LaOFeAs crystallizes in a tetragonal layered
structure, being a member of quaternary oxypnictide family LaOMPn
(M=Cr, Mn, Fe, Co, and Ni, Pn=P and As). Its unit cell, also adopted
as the calculation cell, consists of two formula subunits with eight
atoms limited to the symmetry of $P4/nmm$ space group, in which FeAs
layers and LaO layers are arranged alternating along the c axis. In
the calculations, the experimental lattice parameters $a$=4.03552\AA
, $c$=8.7393\AA were adopted while the two internal coordinates of
La and As atoms within the cell were determined by the energy
minimization. We emphasize the Fe atoms form a 2-dimensional square
lattice with a separation of 2.85\AA. To the end, we checked the
lattice parameters by the energy minimization, and find less than
1\% change so that there are no meaningful changes on the results.

\begin{figure}
\includegraphics[width=7.0cm,height=5.8cm]{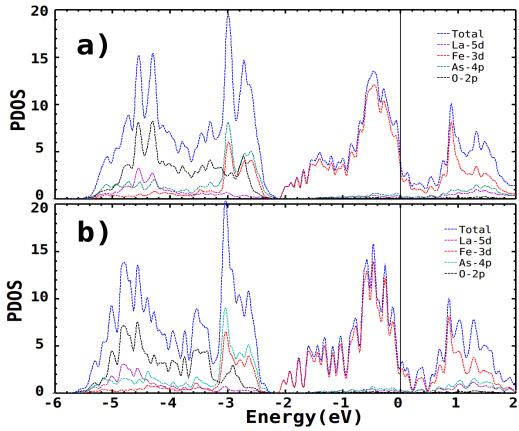}
\caption{(Color online) Calculated atomic orbital-resolved partial
density of states of LaO$_{1-x}$F$_x$FeAs in nonmagnetic state. (a)
parent compound LaOFeAs; (b) LaO$_{0.95}$F$_{0.05}$FeAs (5\%
F-doping). The Fermi energy sets to zero.} \label{figa}
\end{figure}

We first studied the NM state of the compound LaOFeAs, which means
spin degrees were not included in the calculation. Such a study
provides a reference for studying magnetization states, by analyzing
which we can better understand the mechanism underlying the
magnetization if existing. It turns out that our calculated results
on the NM state are similar to those in Ref.
\cite{singh,xu,kotliar}. From Fig. \ref{figa}(a), we see that the
density of states (DOS) may be divided into two parts, i.e. the
lower part (2eV below the Fermi energy) consists of those bands
formed through the bonding between O and La atomic orbitals and Fe
and As atomic orbitals, and the upper part consists basically of the
Fe-$3d$ orbital bands ranging from -2eV to 2eV centered at the Fermi
energy. Further analyzing of the calculation shows that the crystal
field effect upon the Fe-$3d$ orbitals is much weaker than in
transition metal oxides, which is understandable because the
electronegativity of As is much smaller than that of O. Thus all
Fe-$3d$ electrons are expected to play a dominant role in conduction
and related superconductivity if LaOFeAs is nonmagnetic.

\begin{figure}
\includegraphics[width=8.5cm]{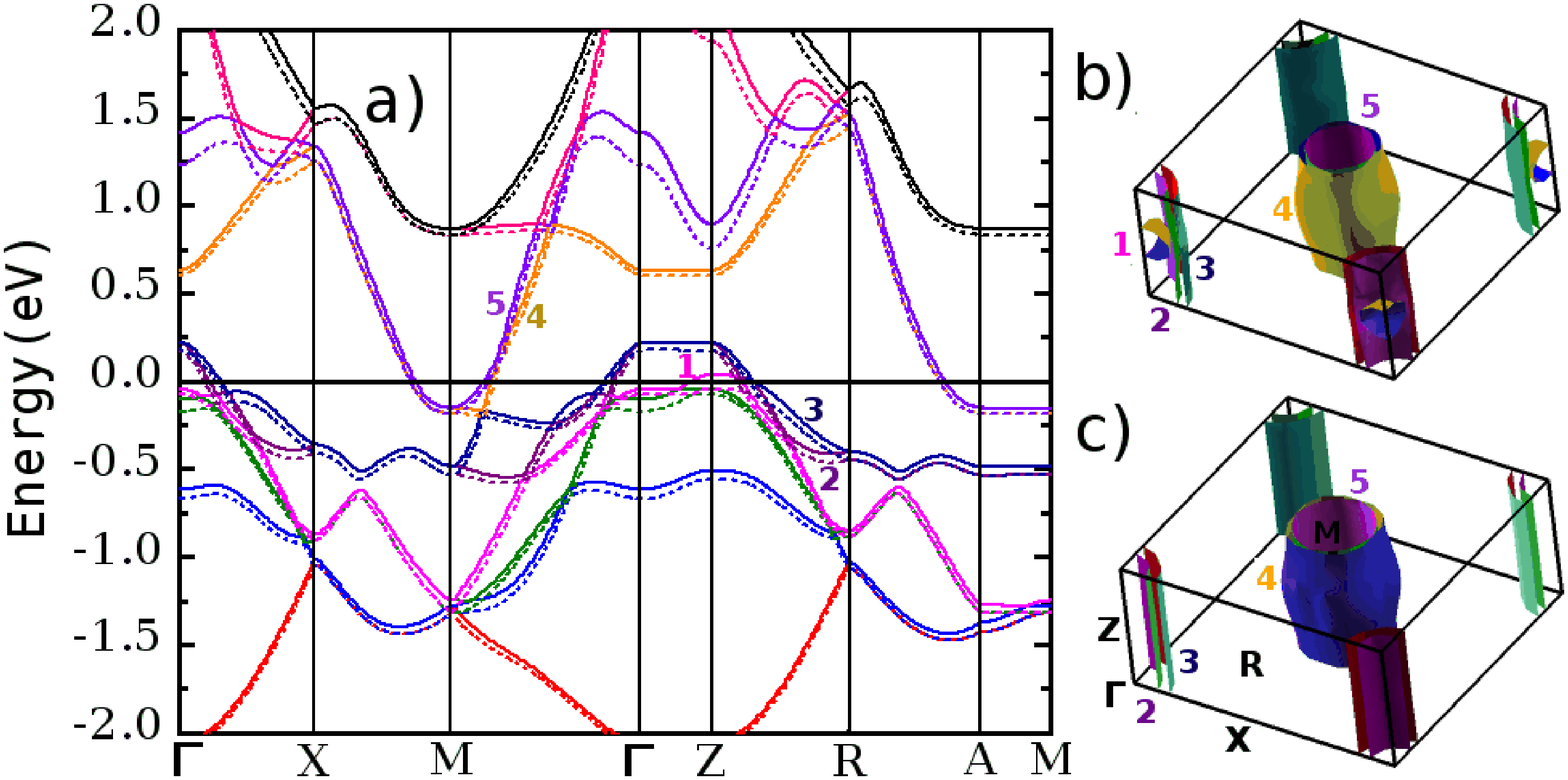}
\caption{(Color online) Calculated nonmagnetic electronic structures
of LaO$_{1-x}$F$_x$FeAs. (a) energy band structures: the solid lines
for the undoping while the dotted lines for the 5\% F-doping (the
Fermi energy sets to zero); (b) undoping: Fermi surface; (c) 5\%
F-doping: Fermi surface.} \label{figc}
\end{figure}

Fig. \ref{figc}(b) shows that the Fermi surface is made up of five
sheets, derived from the five bands crossing the Fermi energy marked
by numbers in Fig. \ref{figc}(a). Among these five sheets, the two
sheets due to two electron bands marked by 4 and 5 are forming two
cylinder-like shapes centered around M-A; and the other three due to
three hole bands marked by 1, 2 and 3 are forming two cylinder-like
shapes centered around $\Gamma$-Z and one pocket around Z,
respectively. As we see, because of nearly no the band dispersion
along (001), the conduction is strongly anisotropic, only happening
on the FeAs layers. The volumes enclosed by these Fermi sheets give
0.28 electrons/cell and equally 0.28 holes/cell, namely $1.97\times
10^{21}$/cm$^3$. The compound LaOFeAs in the NM state is thus a
semimetal with a low carrier density, between normal metals and
semiconductors.

The F-doping effect upon the electronic structures of
LaO$_{1-x}$F$_x$FeAs were studied at $x=3\%,~5\%,~8\%,~10\%$, and
$15\%$ by using virtual crystal calculations. Fig. \ref{figc}(a)
shows that the $5\%$ F-doping just relatively moves the Fermi level
slightly up, i.e. increasing the number of electrons while reducing
the number of holes in the unit cell. This is also clearly shown by
changes of the Fermi surface shown in Fig. \ref{figc}(c), in which
the hole pocket disappears and the two hole cylinders shrink while
the two electron cylinders expand obviously. Such a change is found
up to the 15\% F-doping. Thus the F-doping effect can be basically
considered as doping electrons. On the other hand, Fig. \ref{figa}
(b) shows that the density of states at the Fermi energy heavily
decreases with the F-doping, which suggests that the doping in the
NM state may not favor the superconductivity even though the total
electron carrier density increases. The doping also introduces more
wiggles in the DOS (Fig. \ref{figa} (b)).

\begin{figure}
\includegraphics[width=8.5cm]{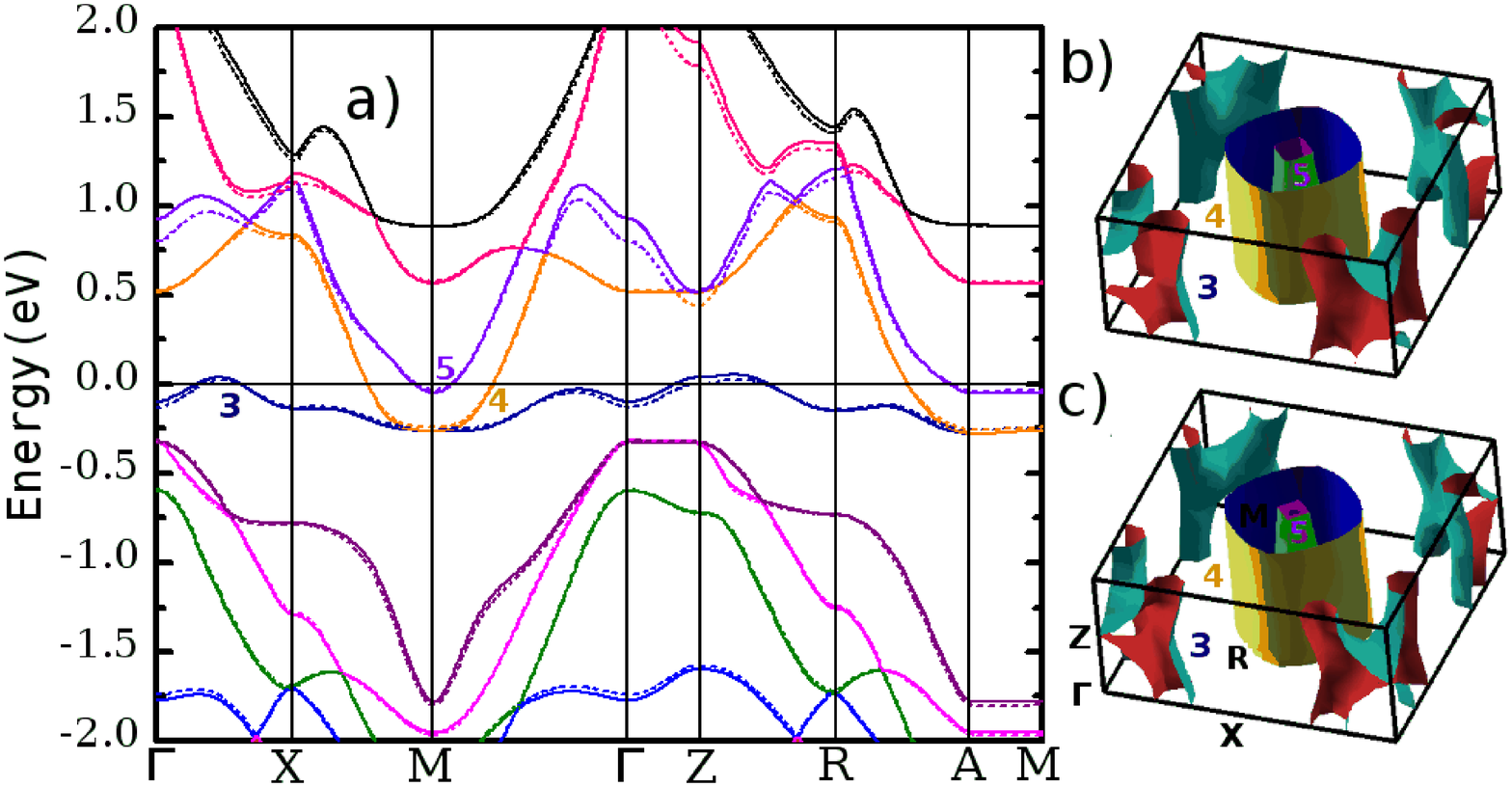}
\caption{(Color online) Calculated antiferromagnetic electronic
structures of LaO$_{1-x}$F$_x$FeAs. (a) energy band structures: the
solid lines for to the undoping while the dotted lines for 5\%
F-doping (the Fermi energy sets to zero); (b) undoping: Fermi
surface; (c) 5\% F-doping: Fermi surface. Note that here only the
spin-up part is shown.} \label{figd}
\end{figure}

We next included spin degrees to study magnetization states of the
compound LaOFeAs. Here we define the spin density polarization as
the ratio of the difference to the sum, between the charge densities
with spin up and spin down. We first broke the spin up-down symmetry
by assigning ferromagnetic moments to the Fe atoms. It turns out no
matter how large to assign the initial moments, the system always
evolves into a NM state without any local moment remaining, namely
back to the initial NM state. LaOFeAs will thus not display
ferromagnetic. We then assigned AFM moments to the two Fe atoms in
the unit cell to break the spin up-down symmetry. We find that when
the initial AFM assignment is with the spin density polarization set
less than 6\%, the system still evolves back into the NM state.
However, when the spin density polarization set larger than 6\%, the
system evolves into a stable AFM state with the energy lowered by
0.16 eV/cell. The internal coordinates of As and La in the unit cell
can be further relaxed, which lowers the energy by extra 0.06
eV/cell. Eventually this AFM state is more stable in energy by 0.22
eV/cell than the NM state. Finally about a 2.38$\mu_B$ magnetic
moment is formed around each Fe atom. We have thus shown that there
exist two stable states, namely the NM state and the AFM state, in
electronic degree configurations for the compound LaOFeAs. Between
the two states, the AFM state is more favorable in energy to be the
ground state and the NM state is then to be a metastable state. We
also did the independent calculation by using the full potential
linearized augmented plane wave (FLAPW) method with the code WIEN2K
package\cite{blaha}, which also indicates the LaOFeAs ground state
is antiferromagnetic.

\begin{figure}
\includegraphics[width=8.0cm]{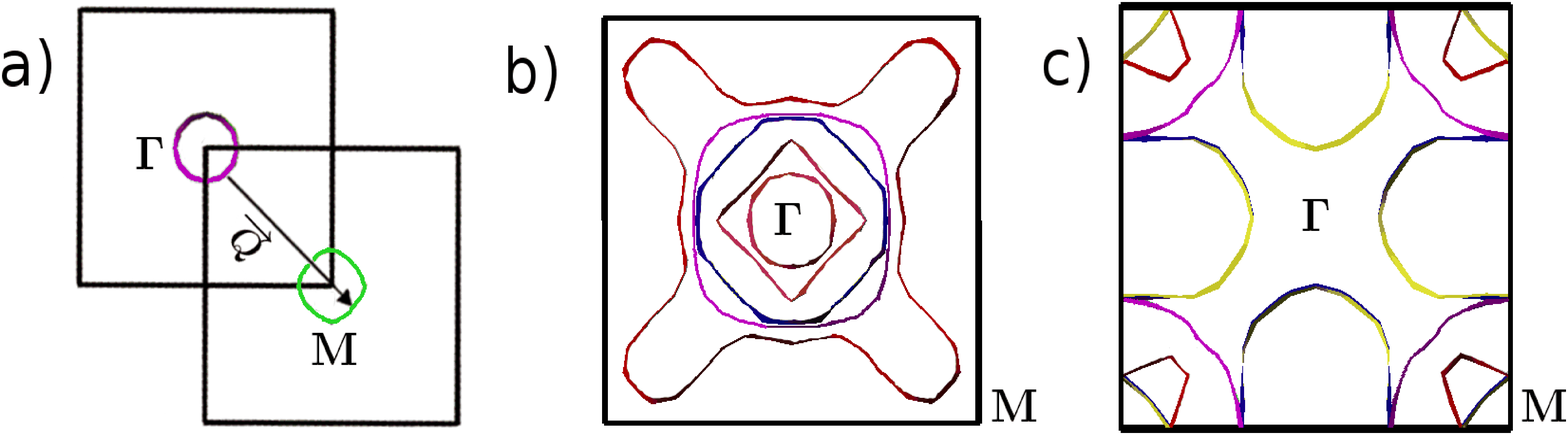}
\caption{(Color online) Calculated nonmagnetic Fermi surface cross
sections through $\Gamma$ and M in (001) plane. (a) LaOFeAs: the
nesting vector is denoted by
$\vec{Q}=(\frac{\pi}{a},\frac{\pi}{a},0)$, between $\Gamma$ and M,
$a$ being the lattice constant; (b)LaOMnAs; (c) LaOCoAs.}
\label{figf}
\end{figure}

\begin{figure}
\includegraphics[width=8.5cm]{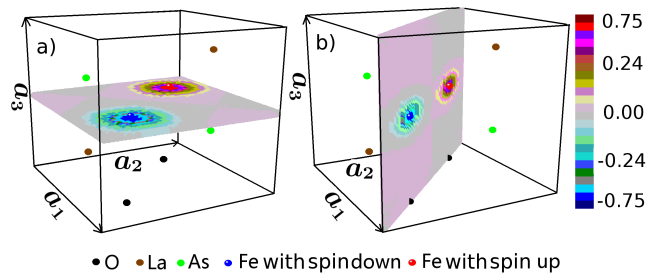}
\caption{(Color online) Calculated antiferromagnetic spin-polarized
charge density distribution in the LaOFeAs unit cell: (a) in the
Fe-Fe square lattice plane; (b) perpendicular to the Fe-Fe square
lattice plane. Here the red part denotes spin up and the blue part
denotes spin down.} \label{fige}
\end{figure}

In comparison with the NM state (Fig. \ref{figc}), there are now
just three bands crossing the Fermi energy (Fig. 3), the previous
two bands 1 and 2 in Fig. \ref{figc}(a) are now pushed down much,
making the hole pocket and one hole cylinder disappear in the AFM
state. Meanwhile, the bands 4 and 5 in Fig. \ref{figc}(a) nearly
emerging around M and A are now splitting, in which the band 4 is
going down while the band 5 is going up in the AFM state. These
changes on the energy band structure overall make the AFM state more
favorable in energy. This can apparently be attributed to the
exchange energy lowering. However, when we compare the Fermi
surfaces between the NM (Fig. 2(b)) and the AFM (Fig. 3(b)) states,
we find that there exists the Fermi surface nesting in the NM Fermi
surface that may induce the AFM Fermi surface, as shown in Fig.
\ref{figf}(a). The nesting vector is
$\vec{Q}=(\frac{\pi}{a},\frac{\pi}{a},0)$ between the hole sheets
and the electron sheets. Fig. \ref{fige} plots the calculated AFM
spin polarized charge density distribution in the LaOFeAs unit cell,
which indicates that the AFM state is a quasi-2-dimensional AFM spin
density wave, oscillating as $\vec{M}\cos(\vec{Q}\cdot\vec{R})$ on
the Fe-Fe square lattice with a wave vector equal to the nesting
vector $\vec{Q}$ ($\vec{R}$ being the lattice site vector). We thus
conclude that the underlying physics to stabilize the AFM state is
the Fermi surface nesting. It turns out that the AFM electronic
structure is quite different from the NM one, especially around the
Fermi energy.

\begin{figure}
\includegraphics[width=7.5cm]{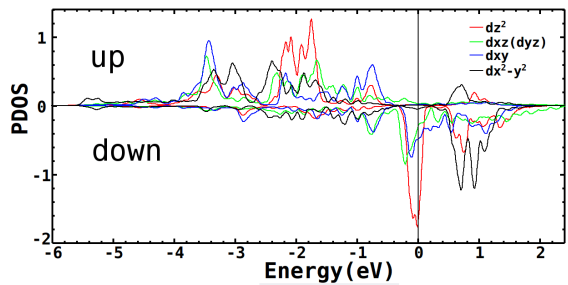}
\caption{(Color online) Calculated antiferromagnetic density of
states projected into the Fe-$3d$ atomic orbitals around one of the
two Fe atoms in the unit cell. Note that the spin-up and the
spin-down is reversed around the other Fe atom.} \label{figg}
\end{figure}

\begin{figure}
\includegraphics[width=7.0cm,height=5.8cm]{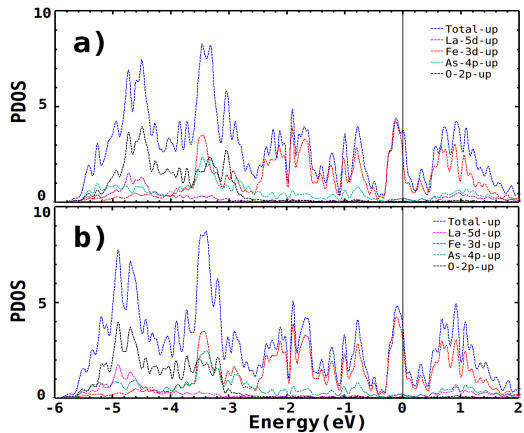}
\caption{(Color online) Calculated atomic orbital-resolved partial
density of states of the compound LaO$_{1-x}$F$_x$FeAs in
antiferromagnetic state. The Fermi energy sets to zero. (a) parent
compound LaOFeAs; (b) LaO$_{0.95}$F$_{0.05}$FeAs (5\% F-doping).
Note that here only the spin-up part is shown.} \label{figb}
\end{figure}

From the calculated AFM density of states shown in Fig. \ref{figb},
we see a gap opens around -0.5eV in contrast to the NM state in Fig.
\ref{figa}(a) because of the Fermi surface nesting. This is also
reflected between the band structures of the NM and AFM states (Fig.
\ref{figc}(a) and Fig. \ref{figd}(a)). Corresponding to this
opening, the more states are pushed down to around -2.0eV.
Meanwhile, the density of states at the Fermi energy becomes larger
(Fig. \ref{figb}(a)), being nearly double (spin up plus down) of
that in the NM case (Fig. \ref{figa}(a)). The projected density of
states in Fig. \ref{figg} further shows that there is basically no
contribution from the Fe-$3d_{x^2-y^2}$ orbitals between $-0.6$eV
and 0.6eV centered at the Fermi energy while the Fe-$3d_{z^2}$ with
$3d_{xz}$, $3d_{yz}$, and $3d_{xy}$ are dominant in this range.
Physically this is because the Fe-$3d_{x^2-y^2}$ orbitals have
formed so strong bond along the nearest Fe-Fe direction that the
derived band is very below the Fermi energy. Further inspection of
the calculations shows that the hole portion in Fig. \ref{figd}(b)
consists mostly of the Fe-$3d_{z^2}$ orbitals with a very small
portion of As-$4p_z$ orbitals while the two electron cylinders
consist of the Fe $3d_{xz}$, $3d_{yz}$, and $3d_{xy}$ orbitals.
Moreover, in the AFM state, the volumes enclosed by the Fermi
surface sheets in Fig. \ref{figd}(b) yield 0.26 electrons/cell
($1.83\times 10^{21}$/cm$^3$) and 0.12 holes/cell ($0.84\times
10^{21}$/cm$^3$), respectively. Thus the parent compound LaOFeAs is
an AFM semimetal with most carriers being electrons. Note that in
the NM state, the carrier densities of electrons and holes are
equal. So we predict that there will be an abrupt change in the Hall
measurement for the compound LaOFeAs from high temperatures to low
temperatures, corresponding to the transition from the NM phase to
the AFM phase.

We also undertook further calculations to check how stable this AFM
state is under the F-doping. Similar to the NM case (Fig.
\ref{figc}(a)), Fig. \ref{figd}(a) shows the 5\% F-doping is also
just rigidly to move the Fermi level slightly up, increasing the
number of electrons while decreasing the number of holes in the unit
cell. The volume enclosed by hole Fermi surface sheet number 3
shrinks obviously in Fig. \ref{figd}(c). We find such a change on
the electronic structure is taken until the 15\% F-doping with
negligible effect on the AFM spin density wave. So the AFM state is
stable against the F-doping. The F-doping effect on the AFM phase is
also basically to dope electrons. Unlike in the NM state, the
density of states at the Fermi energy now decreases not much with
doping (Fig. \ref{figb}(b)).

We notice that the superconductivity takes place in experiment when
the F-doping concentration is between 3\% and 13\% \cite{kamihara},
in which concentration range the compound LaO$_{1-x}$F$_x$FeAs in
normal phase will hold the AFM spin density wave according to our
calculations. Further the itinerant Fe-$3d$ electrons are also
dominant around the Fermi surfaces, so the superconductivity goes
working on the AFM layers with the electron pairing, we suggesting,
mediated by the spin fluctuations.

We also carried out the calculations in order to search for the
trend in line of M=Cr, Mn, Fe, Co, and Ni in LaOMAs and LaOMP. The
calculated results are similar to those reported in Ref. \cite{xu}
except for LaOFeAs that has been reported and analyzed in this
Letter. The scrutiny of the Fermi surface structures further evinces
that there is no more Fermi surface nesting, at least no
commensurate nesting existing, as shown in Fig. \ref{figf} (b) and
(c). The Fermi surface nesting only happens on LaOFeAs, being
underlying mechanism of its AFM, which makes LaOFeAs distinct from
the others in the line.

In conclusion, our studies indicate that the parent compound LaOFeAs
is a quasi-2-dimensional antiferromagnetic semimetal, in which the
antiferromagnetic spin density wave forms on the Fe-Fe square
lattice due to the Fermi surface nesting. This antiferromagnetic
state is robust against the F-doping. This provides a new platform
on which the superconductivity takes place, strongly implying a new
superconductivity mechanism underlying. An abrupt change is
predicted for the Hall measurement due to different carriers between
the nonmagnetic and antiferromagnetic phases.


We wish to thank Prof. Tao Xiang for stimulating and helpful
discussion and Profs. N.L. Wang and J.L. Luo for helpful
conversations. The computing resources are provided by
Super-Computing Center and Institute of Theoretical Physics, Chinese
Academy of Sciences. This work is supported by National Natural
Science Foundation of China (Grant No. 10725419) and by National
Basic Research Program of China (Grant No. 2007CB925004).


\begin{references}

\bibitem{kamihara}Y. Kamihara, T. Watanabe, M. Hirano, and H. Hosono,
J. Am. Chem. Soc. {\bf 130}, 3296 (2008).
\bibitem{muller}J.D. Bednorz and K.A. Muller, Z. Phys. {\bf B64}, 189
(1986); M.K. Wu et. al., Phys. Rev. Lett, {\bf 58}, 908 (1987).
\bibitem{singh}D.J. Singh and M.H. Du, cond-mat/0803.0429v1.
\bibitem{xu}G. Xu, W. Ming, Y. Yao, X. Dai, and Z. Fang, cond-mat/0803.1282v1.
\bibitem{kotliar}K. Haule, J.H. Shim, and G. Kotliar, cond-mat/0803.1279v1.
\bibitem{moruzzi}V.L. Moruzzi, P.M. Marcus, K. Schwarz, and P. Mohn,
Phys. Rev. B {\bf 34}, 1784 (1986); C.S. Wang, B.M. Klein, and H.
Krakauer, Phys. Rev. Lett. {\bf 54}, 1852 (1985); M. Korling and J.
Ergon, Phys. Rev. B {\bf 54}, R8293 (1996); Y.M. Zhou, D.S. Wang,
and Y. Kawazoe, Phys. Rev. B {\bf 59}, 8387 (1999).
\bibitem{pwscf}P. Giannozzi et al., http://www.quantum-espresso.org.
\bibitem{pbe}J. P. Perdew, K. Burke, and M. Erznerhof,
Phys. Rev. Lett. {\bf 77}, 3865 (1996).
\bibitem{vanderbilt}D. Vanderbilt, Phys. Rev. B {\bf 41}, 7892 (1990).
\bibitem{blaha}P. Blaha, {\em et al.}, WIEN2K, An Augmented Plane
Wave + Local Orbits Program for Calculated Crystal Properties (K.
Schwarz, TU Wien, Austria, 2001).

\end{references}
\end{document}